\documentclass[aps,prd,nofootinbib,superscriptaddress,floatfix,nofootinbib,amsmath,amssymb]
{revtex4}
\usepackage[dvips]{graphicx}

\begin{document}

\title{Revisit of Cosmic Age Problem}

\author{Shuang Wang}
\email{swang@mail.ustc.edu.cn}
\affiliation{Department of Modern Physics, University of Science and Technology of China, Hefei 230026, China}
\affiliation{Institute of Theoretical Physics, Chinese Academy of Sciences, Beijing 100080, China}

\author{Xiao-Dong Li}
\email{renzhe@mail.ustc.edu.cn}
\affiliation{Interdisciplinary Center for Theoretical Study, University of Science and Technology of China, Hefei 230026, China}
\affiliation{Institute of Theoretical Physics, Chinese Academy of Sciences, Beijing 100080, China}

\author{Miao Li}
\email{mli@itp.ac.cn}
\affiliation{Institute of Theoretical Physics, Chinese Academy of Sciences, Beijing 100080, China}
\affiliation{Kavli Institute for Theoretical Physics China, Chinese Academy of Sciences, Beijing 100080, China}
\affiliation{Key Laboratory of Frontiers in Theoretical Physics, Chinese Academy of Sciences, Beijing 100080, China}

\begin{abstract}

We investigate the cosmic age problem associated with 9 extremely old globular clusters in M31 galaxy and 1 very old high-$z$ quasar APM 08279 + 5255 at $z=3.91$.
These 9 globular clusters have not been used to study the cosmic age problem in the previous literature.
By evaluating the age of the universe in the $\Lambda$CDM model with the observational constraints from the SNIa, the BAO, the CMB, and the independent $H_0$ measurements,
we find that the existence of 5 globular clusters and 1 high-$z$ quasar are in tension (over 2$\sigma$ confidence level) with the current cosmological observations.
So if the age estimates of these objects are correct, the cosmic age puzzle still remains in the standard cosmology.
Moreover, we extend our investigations to the cases of the interacting dark energy models.
It is found that although the introduction of the interaction between dark sectors can give a larger cosmic age,
the interacting dark energy models still have difficulty to pass the cosmic age test.

\end{abstract}

\maketitle

\section{Introduction}\label{sec:intro}

Before the great discovery that our universe is undergoing an accelerated expansion \cite{Riess,Perlmutter},
the most popular cosmological model was the SCDM model (i.e., a flat universe with the present fractional matter density $\Omega_{m}=1$).
However, the SCDM model is always plagued by a longstanding puzzle:
in this model the present age of the universe is $t_0=\frac{2}{3 H_0}\simeq 9$ Gyr ($H_0$ denotes the Hubble constant),
while astronomers have already discovered that many objects are older than 10 Gyr \cite{chaboyer}.
This so-called cosmic age problem becomes more acute if one considers the age of the universe at a high redshift.
For instance, a 3.5 Gyr-old galaxy 53W091 at redshift $z=1.55$ and a 4 Gyr-old galaxy 53W069 at $z=1.43$ are more difficult to accommodate in the SCDM model \cite{Dunlop}.
Along with the discovery of accelerated expansion of our universe and the return of the cosmological constant $\Lambda$,
the age problem has been greatly alleviated.
The 7-year WMAP observations \cite{WMAP7} tell us that in the standard cosmological model (i.e. the $\Lambda$CDM model) the present cosmic age is $t_0=13.75\pm0.11$ Gyr.
Besides, it is shown that the $\Lambda$CDM model can also easily accommodate galaxies 53W091 and 53W069 \cite{Lima1}.
Therefore, it is widely believed that the cosmic age problem is a ``smoking-gun" evidence for the standard cosmological model,
and many people believe that the cosmic age puzzle does not exist in the $\Lambda$CDM model.
However, we still want to raise the question: Is the cosmic age problem really removed by the $\Lambda$CDM model?

In a recent paper \cite{M31_1}, by comparing photometric data acquired from the Beijing-Arizona-Taiwan-Connecticut (BATC) system
with up-to-date theoretical synthesis models,
the ages of 35 globular clusters (GCs) in M31 galaxy were estimated by Ma et al.
Soon after, the ages of other 104 GCs in M31 galaxy were also obtained by the same research group \cite{M31_2}.
From their results, we find that 9 extremely old GCs are older than the present cosmic age ($t_0=13.75$ Gyr) predicted by the 7-year WMAP observations (see Table \ref{01}).
To our present knowledge, these 9 GCs have not been used to study the cosmic age problem in the previous literature;
so it is interesting to explore  implications of these 9 old GCs for the standard cosmology.
In addition, the existence of high-$z$ quasar APM 08279 + 5255 at $z = 3.91$ \cite{Hasinger,Lima2} still remains a mystery \cite{Lima2,Jain,Pires,BW,swang1}.
Using the maximum likelihood values of the 7-year WMAP observations $\Omega_{m} =0.272$ and $h=0.704$ ($h$ is the reduced Hubble parameter) \cite{WMAP7},
it is seen that the $\Lambda$CDM model can only give a cosmic age $t=1.63$ Gyr at $z = 3.91$, while the lower limit of the quasar's age is $1.8$ Gyr \cite{Lima2}.
In a previous paper \cite{swang1}, Wang and Zhang demonstrated that by simply introducing the dark energy (DE) alone we cannot remove the high-$z$ age problem,
and suggested that the introduction of interaction between dark sectors may be helpful to alleviate the cosmic age problem.
But in \cite{swang1}, when evaluating the age of the universe in the interacting DE models,
the parameters describing interaction strength were chosen arbitrarily.
A consistent quantitative analysis should use the maximum likelihood method to determine the corresponding model parameters from the current cosmological observations.
Then using the obtained results one can analyze the consistency with the cosmic age data.
This analysis will be done in the present work.

\begin{table}
\caption{The ages of 9 extremely old Globular Clusters in M31
Galaxy. Notice that the present cosmic age predicted by the 7-year
WMAP observations is $t_0=13.75\pm0.11$ Gyr.}
\begin{center}
\label{01}
\begin{tabular}{|c|c|c|}
  \hline
  Object & Age(Gyr) & Reference \\
  \hline
  B024 & $15.25\pm 0.75$ & Table 5 of \cite{M31_2} \\
  \hline
  B050 & $16.00\pm 0.30$  & Table 5 of \cite{M31_1} \\
  \hline
  B129 & $15.10\pm 0.70$  & Table 5 of \cite{M31_1} \\
  \hline
  B144D & $14.36\pm 0.95$  & Table 5 of \cite{M31_2} \\
  \hline
  B239 & $14.50\pm 2.05$  & Table 5 of \cite{M31_1} \\
  \hline
  B260 & $14.30\pm 0.50$  & Table 5 of \cite{M31_2} \\
  \hline
  B297D & $15.18\pm 0.85$  & Table 5 of \cite{M31_2} \\
  \hline
  B383 & $13.99\pm 1.05$  & Table 5 of \cite{M31_2} \\
  \hline
  B495 & $14.54\pm 0.55$  & Table 5 of \cite{M31_2} \\
  \hline
\end{tabular}
\end{center}
\end{table}

This paper is organized as follows.
In section II, we evaluate the age of the universe in the $\Lambda$CDM model with constraints from
the Type Ia supernovae (SNIa), the Baryon Acoustic Oscillations (BAO), the Cosmic Microwave Background (CMB), and the independent $H_0$ measurements,
then use the obtained results to analyze the consistency with the age data listed above.
In section III, we extend our investigation to the cases of three interacting dark energy models.
We summarize in Section IV.
Appendix A supplies the details of how to utilize the cosmological observations to constrain DE models.
We assume today's scale factor $a_{0}=1$, so the redshift $z=a^{-1}-1$,
the subscript ``0'' always indicates the present value of the corresponding quantity.

\

\section{Cosmic Age Problem in The $\Lambda$CDM model}\label{sec:LCDM}

The age $t(z)$ of a flat universe at redshift $z$ is \cite{Jain}
\begin{equation} \label{01}
t(z)=\int_z^\infty\frac{d\tilde{z}}{(1+\tilde{z})H(\tilde{z})},
\end{equation}
where $H(z)$ is the Hubble parameter, given by the Friedman
equation,
\begin{equation} \label{02}
3M_{Pl}^2H^2=\rho_m+\rho_{de}+\rho_{r},
\end{equation}
$M_{Pl}\equiv 1/\sqrt{8\pi G}$ is the reduced Planck mass, $\rho_m$,
$\rho_{de}$, and $\rho_{r}$ are the energy density of matter, dark
energy, and radiation, respectively. The reduced Hubble parameter is
$E(z)\equiv H(z)/H_{0}$. Once $E(z)$ is given, the cosmic age $t(z)$
at any redshift $z$ can be evaluated. For $\Lambda$CDM model,
\begin{equation} \label{03}
E(z)=\sqrt{\Omega_{r}(1+z)^4+\Omega_m(1+z)^3+1-\Omega_m-\Omega_{r}},
\end{equation}
where $\Omega_{r}$ is the present fractional radiation density given
by \cite{WMAP7},
\begin{equation}
\Omega_{r}=\Omega_{\gamma}(1+0.2271N_{eff}),\ \ \
\Omega_{\gamma}=2.469\times10^{-5}h^{-2},\ \ \ N_{eff}=3.04.
\end{equation}
Note that in a flat universe the present fractional DE density is
$\Omega_{de}=1-\Omega_m-\Omega_{r}$. Since we also consider effects
of radiation component, there are two model parameters, $\Omega_{m}$
and $h$, need to be determined.

To constrain DE models by using cosmological observations, we will
employ the $\chi^2$ statistics \cite{mli}. For a physical quantity
$\xi$ with an experimentally measured value $\xi_{obs}$, a standard
deviation $\sigma_{\xi}$, and a theoretically predicted value
$\xi_{th}$, the $\chi^2$ value is given by
\begin{equation} \label{eq:chi2_xi}
\chi_{\xi}^2=\frac{\left(\xi_{th}-\xi_{obs}\right)^2}{\sigma_{\xi}^2}.
\end{equation}
The total $\chi^2$ is the sum of all $\chi_{\xi}^2$s, i.e.,
\begin{equation} \label{eq:chi2}
\chi^2=\sum_{\xi}\chi_{\xi}^2.
\end{equation}
The model parameters yielding a minimal $\chi^2$ is favored by
observations. The observational data used in this paper include the
Constitution SNIa sample \cite{SN09}, the BAO data from the SDSS
Data Release 7 (DR7) galaxy sample \cite{BAO2}, the CMB measurements
given by the 7-year WMAP observations \cite{WMAP7}, and the
independent $H_0$ measurements from HST \cite{H0} (see Appendix A
for details). With these observational data, the cosmological
constraints on the $\Lambda$CDM model are obtained:
$\Omega_{m}=0.275_{-0.024}^{+0.026}$ and $h=0.704_{-0.019}^{+0.020}$
at 1$\sigma$ confidence level (CL), while
$\Omega_{m}=0.275_{-0.038}^{+0.043}$ and $h=0.704_{-0.031}^{+0.033}$
at 2$\sigma$ CL.

\begin{figure}
\includegraphics[width=8.5cm]{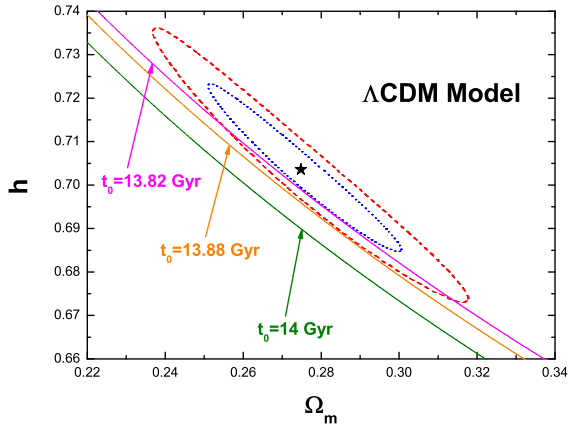}
\caption{Cosmic age problem at $z=0$ in the $\Lambda$CDM model. The
dotted elliptical line, the dashed elliptical line, and the central
star symbol denote the 1$\sigma$ confidence region, the 2$\sigma$
confidence region, and the best-fit point of the $\Lambda$CDM model
in the $\Omega_{m}$-$h$ plane, respectively. Three solid diagonal
lines, from top to bottom, represent the constraints of $13.82$ Gyr,
$13.88$ Gyr, and $14$ Gyr at $z=0$, respectively. Notice that the
top diagonal line tangents to the 1$\sigma$ confidence region of the
$\Lambda$CDM model, while the middle diagonal line tangents to the
2$\sigma$ confidence region of the $\Lambda$CDM model. For a given
diagonal line, the area below this diagonal line corresponds to a
larger cosmic age. The $\Lambda$CDM model predicts a lower cosmic
age than the 1$\sigma$ lower limits of 5 GCs' age. \label{fig1}}
\end{figure}

Now we discuss the cosmic age problem associated with those 9 extremely old GCs in M31 galaxy.
The results are given in Fig.\ref{fig1}.
Our method is as follows.
Firstly, since in a flat universe the age problem is mainly constrained by observational data of ($\Omega_{m}$, $h$),
by using the current cosmological observations (SNIa+BAO+CMB+$H_0$),
we plot the 1$\sigma$ and the 2$\sigma$ confidence regions of the $\Lambda$CDM model in the $\Omega_{m}$-$h$ plane.
Secondly, since a specific cosmic age corresponds to a specific diagonal line in the $\Omega_{m}$-$h$ plane
and a larger cosmic age corresponds to a lower diagonal line,
by searching the diagonal line that tangents to the 2$\sigma$ confidence region of the $\Lambda$CDM model,
we obtain the theoretical upper limit of present cosmic age (at 2$\sigma$ CL) predicted by the $\Lambda$CDM model.
As seen in the figure, this theoretical upper limit is $t_0\leq 13.88$ Gyr.
Notice that this upper limit is only given by the combined SNIa+BAO+CMB+$H_0$ data, and has nothing to do with the age data of those GCs.
Thirdly, we compare this theoretical upper limit of present cosmic age with the observational lower limits of those old GCs' ages listed in Table \ref{01},
and find that the 1$\sigma$ lower limits of 5 GCs' age (including B024 with a lower limit 14.50 Gyr, B050 with a lower limit 15.70 Gyr,
B129 with a lower limit 14.40 Gyr, B297D with a lower limit 14.33 Gyr, and B495 with a lower limit 13.99 Gyr)
are even larger than 13.88 Gyr.
So the existence of these 5 GCs are in tension (over 2 $\sigma$ CL) with the current cosmological observations.
This means that if the age estimates of these 5 GCs \cite{M31_1,M31_2} are correct, the cosmic age puzzle still remains in the standard cosmology.
It should be mentioned that during this analysis, only 9 most discrepant GCs are used to quote statistical discrepancy.
One may wonder what the discrepancy is when all objects are taken into account.
By treating all those 139 GCs in M31 galaxy as data points and utilizing the $\chi^2$ statistics,
this issue is also checked.
We find that the $\Lambda$CDM model can accommodate 127 GCs,
and the statistical discrepancy disappears when all objects are included.

\begin{figure}
\includegraphics[width=8.5cm]{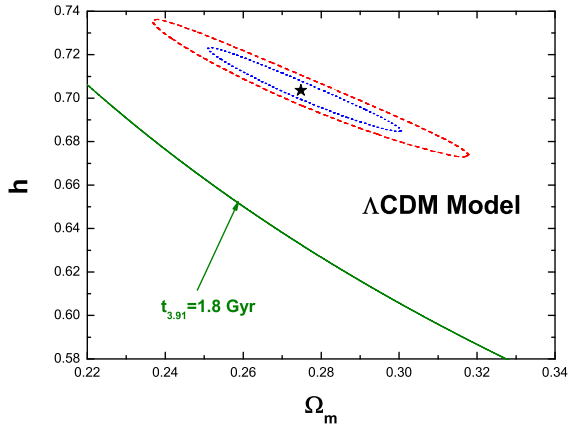}
\caption{High-$z$ cosmic age problem at $z=3.91$ in the $\Lambda$CDM
model. The dotted elliptical line, the dashed elliptical line, and
the central star symbol denote the 1$\sigma$ confidence region, the
2$\sigma$ confidence region, and the best-fit point of the
$\Lambda$CDM model in the $\Omega_{m}$-$h$ plane, respectively. The
solid diagonal line represents the lower limit of the quasar's age
(i.e. $1.8$ Gyr) at $z=3.91$. The area below this diagonal line
corresponds to a cosmic age larger than $1.8$ Gyr. The $\Lambda$CDM
model predicts a lower cosmic age than the 1$\sigma$ lower limit of
the old quasar's age. \label{fig2}}
\end{figure}

Next we turn to the high-$z$ cosmic age problem associated with quasar APM 08279 + 5255 at $z=3.91$.
The age of this quasar had been estimated by studying its chemical evolution.
Based on the evolution of Fe/O ratio from the X-ray observation, Ref. \cite{Hasinger} gave a rough estimate $t_{QSO}=(2.0 \sim 3.0)$ Gyr at z = 3.91.
Soon after, by using a detailed chemodynamical model for the evolution of spheroid, Friaca et al. \cite{Lima2} obtained $t_{QSO}=2.1\pm 0.3$ Gyr at the same redshift.
The high-$z$ cosmic age problem at $z=3.91$ in the $\Lambda$CDM model is demonstrated in Fig.\ref{fig2}.
The method of analysis is same as that used in Fig.\ref{fig1}.
Notice that the solid diagonal line represents the 1$\sigma$ lower limit of the quasar's age (i.e. $1.8$ Gyr) at $z=3.91$.
Our calculations show that the existence of quasar 08279 + 5255 is evidently inconsistent with the current cosmological observations,
and the discrepancy is over 7$\sigma$ CL.

\

\section{Cosmic Age Problem in The Interacting DE Models}\label{sec:IDE}

As seen in the previous section, the $\Lambda$CDM model has
difficulty to pass the cosmic age test. One may ask how other DE
models fare. In a previous paper \cite{swang1}, Wang and Zhang demonstrated that by simply introducing
the DE alone we cannot remove the high-$z$ age problem, and
suggested that the introduction of interaction between dark sectors
may be helpful to alleviate the cosmic age problem. So in this work,
we also investigate the cosmic age problem in the interacting DE
models. As examples, three kinds of interacting DE models are
considered here. In the first subsection, we introduce these 3
interacting DE models, and determine their model parameters by using
cosmological observations. Then in the second subsection, we
evaluate the age of the universe in these 3 models, and use the
obtained results to analyze the consistency with the age data listed
above.

\subsection{Cosmological Constraints on Three Interacting DE Models}

The universe is filled with three classes of major energy components:
matter (including baryons and dark matter), dark energy, and
radiation. After introducing interaction between dark sectors, the
dynamical evolutions of these 3 components satisfy
\begin{eqnarray}
\label{eq:CEt1}&& \dot \rho_m+3H\rho_m=Q\ \ \\
\label{eq:CEt2}&& \dot \rho_{de}+3H(\rho_{de}+p_{de})=-Q\ \ \\
\label{eq:CEt3}&& \dot \rho_r+4H\rho_r=0
\end{eqnarray}
where the over dot denotes the derivative with respect to the cosmic
time $t$, $Q$ denotes the phenomenological interaction term, and
$p_{de}$ is the pressure of DE given by the specific model.

We consider a simple scalar DE model with Lagrangian \cite{Peebles}
\begin{equation} \label{A1}
L=\frac{1}{2}\partial^{\mu}\phi\partial_{\mu}\phi-V(\phi).
\end{equation}
Assuming the potential energy $ V(\phi)$ be dominant, the energy
density and the pressure of DE are
\begin{equation} \label{A3}
\rho_{de}=-p_{de} \simeq V.
\end{equation}
This means that the equation of state of DE $w\equiv
p_{de}/\rho_{de}=-1$. Noting that $a=\frac{1}{1+z}$, $H=\frac{\dot
a}{a}$, we have
\begin{equation}
\frac{d}{dz}=\frac{d}{dt}\frac{dt}{da}\frac{da}{dz}=-\frac{1}{H(1+z)}\frac{d}{dt}.
\end{equation}
Then Eqs.(\ref{eq:CEt1}-\ref{eq:CEt3}) can be rewritten as
\begin{eqnarray}
\label{eq:CEz1}&&(1+z)\frac{d\rho_m}{dz}-3\rho_m=-Q/H\ \ \\
\label{eq:CEz2}&&(1+z)\frac{d\rho_{de}}{dz}=Q/H\ \ \\
\label{eq:CEz3}&&(1+z)\frac{d\rho_r}{dz}-4\rho_r=0
\end{eqnarray}
Eq.(\ref{eq:CEz3}) has a general solution $\rho_r=\rho_{r0}(1+z)^4$,
while the solutions of Eqs.(\ref{eq:CEz1}) and (\ref{eq:CEz2})
depend on the specific form of $Q$.

Owing to lack of the knowledge of microscopic origin of the
interaction, here we just follow \cite{mli2} and consider three classes of interaction terms
\begin{eqnarray}
\label{eq:Int1}&&Q_1=3\alpha H\rho_m\\
\label{eq:Int2}&&Q_2=3\alpha H\rho_{de}\\
\label{eq:Int3}&&Q_3=3\alpha H(\rho_m+\rho_{de})
\end{eqnarray}
where $\alpha$ is a model parameter denoting the strength of
interaction. For simplicity, we will denote these 3 models by
I$\Lambda$CDM1, I$\Lambda$CDM2, and I$\Lambda$CDM3, respectively.
Notice that all these interacting DE models have 3 model parameters,
$\Omega_{m}$, $\alpha$, and $h$.

First we discuss the I$\Lambda$CDM1 model. Substituting $Q_1=3\alpha
H\rho_m$ into Eq.(\ref{eq:CEz1}), we have
\begin{equation} \label{eq:mILCDM1}
\rho_m=\rho_{m0}(1+z)^{3(1-\alpha)}.
\end{equation}
Substituting Eq.(\ref{eq:mILCDM1}) into Eq.(\ref{eq:CEz2}) and using
the initial condition $\rho_{de}(z=0)=\rho_{de0}$, we get
\begin{equation} \label{eq:deILCDM1}
\rho_{de}=\frac{\alpha
\rho_{m0}}{1-\alpha}(1+z)^{3(1-\alpha)}-\frac{\alpha\rho_{m0}}{1-\alpha}+\rho_{de0}.
\end{equation}
Combining Eqs.(\ref{eq:mILCDM1}), (\ref{eq:deILCDM1}), and
(\ref{02}), we obtain the reduced Hubble parameter of the
I$\Lambda$CDM1 model
\begin{equation} \label{eq:EzILCDM1}
E(z)=\sqrt{\Omega_{r}(1+z)^4+\frac{\Omega_m}{1-\alpha}(1+z)^{3(1-\alpha)}+\big(1-\Omega_{r}-\frac{\Omega_m}{1-\alpha}\big)}.
\end{equation}
Based on the $\chi^2$ statistics method and observational data, the
model parameters of the I$\Lambda$CDM1 model are determined as
$\Omega_{m}=0.279_{-0.027}^{+0.028}$,
$\alpha=(-1.20\times10^{-3})_{-3.14\times10^{-3}}^{+3.45\times10^{-3}}$,
and $h=0.705_{-0.019}^{+0.020}$ at 1$\sigma$ CL, while
$\Omega_{m}=0.279_{-0.043}^{+0.047}$,
$\alpha=(-1.20\times10^{-3})_{-5.01\times10^{-3}}^{+5.78\times10^{-3}}$,
and $h=0.705_{-0.031}^{+0.033}$ at 2$\sigma$ CL.

We now turn to the I$\Lambda$CDM2 model. Substituting $Q_2=3\alpha
H\rho_{de}$ into Eq.(\ref{eq:CEz2}), we have
\begin{equation} \label{eq:deILCDM2}
\rho_{de}=\rho_{de0}(1+z)^{3\alpha}.
\end{equation}
Substituting Eq.(\ref{eq:deILCDM2}) into Eq.(\ref{eq:CEz1}) and
using the initial condition $\rho_{m}(z=0)=\rho_{m0}$, we get
\begin{equation} \label{eq:mILCDM2}
\rho_{m}=\rho_{m0}(1+z)^3-\frac{\alpha\rho_{de0}}{1-\alpha}(1+z)^3+\frac{\alpha\rho_{de0}}{1-\alpha}(1+z)^{3\alpha}.
\end{equation}
Combining Eqs.(\ref{eq:deILCDM2}), (\ref{eq:mILCDM2}), and
(\ref{02}), we obtain the reduced Hubble parameter of I$\Lambda$CDM2
model
\begin{equation} \label{eq:EzILCDM2}
E(z)=\sqrt{\Omega_{r}(1+z)^4+\frac{\Omega_{m}-\alpha+\alpha\Omega_{r}}{1-\alpha}(1+z)^3+\frac{1-\Omega_m-\Omega_{r}}{1-\alpha}(1+z)^{3\alpha}}.
\end{equation}
The model parameters of the I$\Lambda$CDM2 model are determined as
$\Omega_{m}=0.276_{-0.026}^{+0.028}$,
$\alpha=(-1.43\times10^{-3})_{-1.58\times10^{-2}}^{+1.34\times10^{-2}}$,
and $h=0.702_{-0.024}^{+0.026}$ at 1$\sigma$ CL, while
$\Omega_{m}=0.276_{-0.042}^{+0.046}$,
$\alpha=(-1.43\times10^{-3})_{-2.75\times10^{-2}}^{+2.10\times10^{-2}}$,
and $h=0.702_{-0.038}^{+0.043}$ at 2$\sigma$ CL.

Next we consider the I$\Lambda$CDM3 model. Substituting $Q_3=3\alpha
H(\rho_m+\rho_{de})$ into Eqs.(\ref{eq:CEz1}) and (\ref{eq:CEz2}),
we have
\begin{eqnarray}
\label{eq:ILCDM3_1}&&(1+z)\frac{d\rho_m}{dz}=3(1-\alpha)\rho_m-3\alpha\rho_{de}\\
\label{eq:ILCDM3_2}&&(1+z)\frac{d\rho_{de}}{dz}=3\alpha(\rho_m+\rho_{de})
\end{eqnarray}
Utilizing the initial conditions $\rho_{m}(z=0)=\rho_{m0}$ and
$\rho_{de}(z=0)=\rho_{de0}$, the differential equations
(\ref{eq:ILCDM3_1}) and (\ref{eq:ILCDM3_2}) are solved as
\begin{equation}
\label{eq:ILCDM3_rho}\rho_m=C_1(1+z)^{k_1}+C_2(1+z)^{k_2},\ \ \ \
\rho_{de}=C_1\frac{3-3\alpha-k_1}{3\alpha}(1+z)^{k_1}+C_2\frac{3-3\alpha-k_2}{3\alpha}(1+z)^{k_2},
\end{equation}
where
\begin{equation} \label{eq:ILCDM3_k1k2}
k_1=\frac{3}{2}(1+\sqrt{1-4\alpha}),\ \ \ \
k_2=\frac{3}{2}(1-\sqrt{1-4\alpha}),
\end{equation}
and
\begin{equation}
C_1=\frac{(3-3\alpha-k_2)\rho_{m0}-3\alpha\rho_{de0}}{k_1-k_2},\ \ \
C_2=\frac{(3-3\alpha-k_1)\rho_{m0}-3\alpha\rho_{de0}}{k_2-k_1}.
\end{equation}
From the Friedman equation, we obtain the reduced Hubble parameter
of the I$\Lambda$CDM3 model
\begin{equation} \label{eq:EzILCDM3}
E(z)=\sqrt{\Omega_{r}(1+z)^4+\tilde{C}_1\frac{3-k_1}{3\alpha}(1+z)^{k_1}+\tilde{C}_2\frac{3-k_2}{3\alpha}(1+z)^{k_2}}
\end{equation}
where
\begin{equation}
\tilde{C}_1=\frac{(3-3\alpha-k_2)\Omega_{m}-3\alpha(1-\Omega_{m}-\Omega_{r})}{k_1-k_2},\
\ \
\tilde{C}_2=\frac{(3-3\alpha-k_1)\Omega_{m}-3\alpha(1-\Omega_{m}-\Omega_{r})}{k_2-k_1}.
\end{equation}
Parameters of the I$\Lambda$CDM3 model are determined as
$\Omega_{m}=0.279_{-0.027}^{+0.028}$,
$\alpha=(-8.98\times10^{-4})_{-2.72\times10^{-3}}^{+2.86\times10^{-3}}$,
and $h=0.704_{-0.019}^{+0.019}$ at 1$\sigma$ CL, while
$\Omega_{m}=0.279_{-0.043}^{+0.047}$,
$\alpha=(-8.98\times10^{-4})_{-4.42\times10^{-3}}^{+4.70\times10^{-3}}$,
and $h=0.704_{-0.030}^{+0.032}$ at 2$\sigma$ CL.

\subsection{Age Problem of Three Interacting DE Models}

In this subsection, we evaluate the age of the universe in these 3 interacting DE models,
and analyze the consistency with the age data listed above.
Since in a flat universe the age problem is mainly constrained by the observational data of ($\Omega_{m}$, $h$),
our results are given in the $\Omega_{m}$-$h$ plane.
The method of analysis is same as that used in section II.

\begin{figure}
\includegraphics[scale=0.6, angle=0]{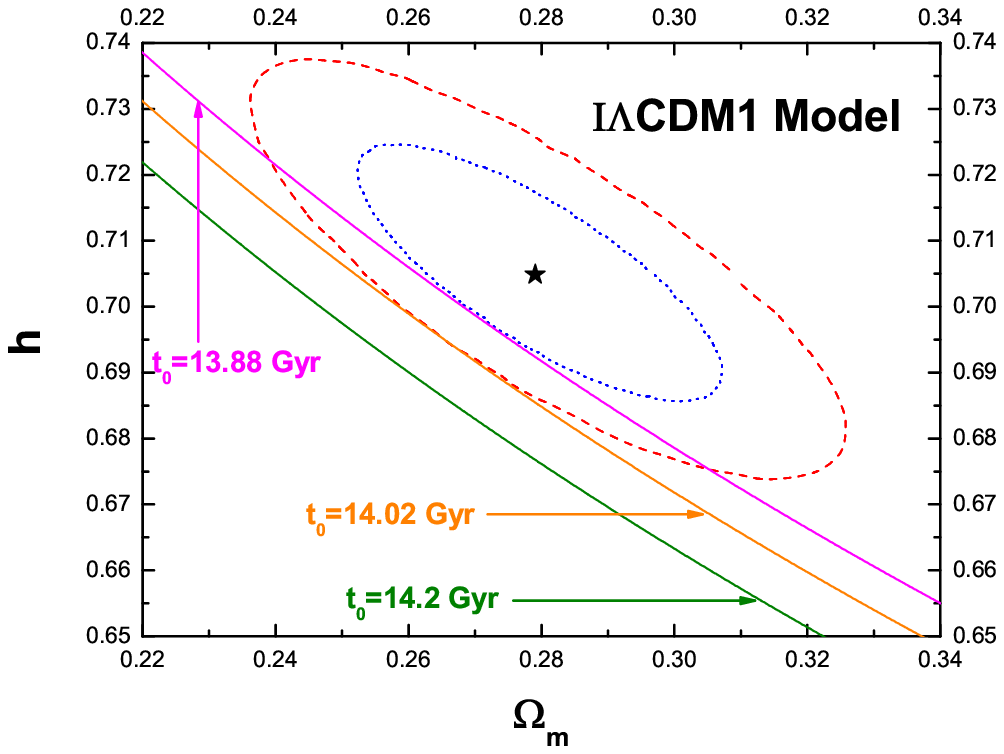}
\includegraphics[scale=0.6, angle=0]{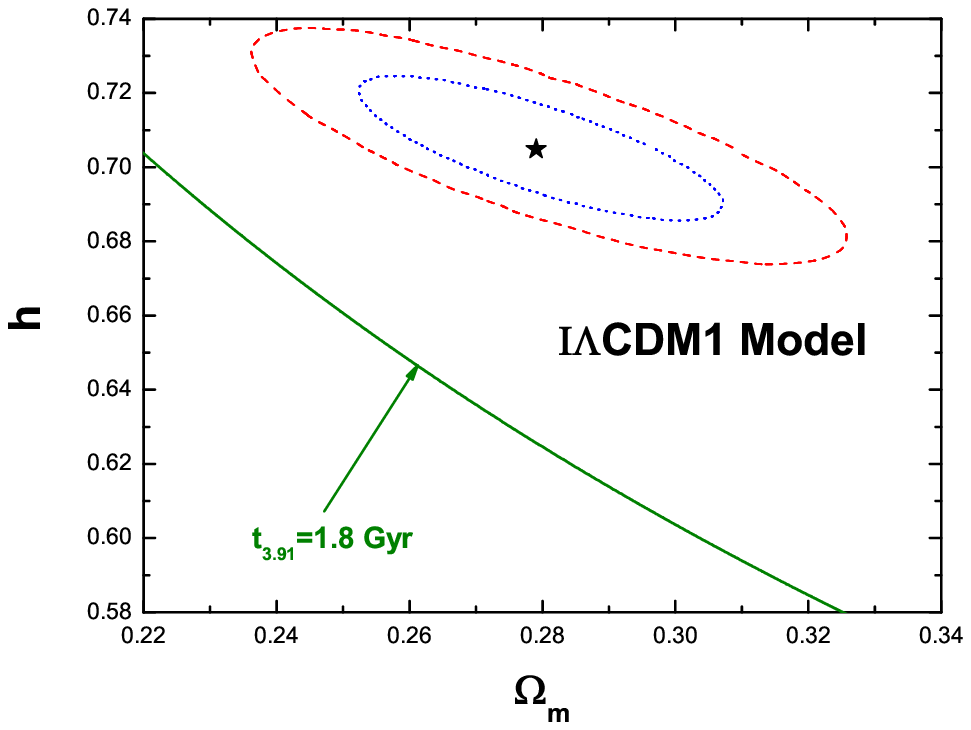}
\caption{\label{fig3} Cosmic age problem in I$\Lambda$CDM1 model.
The dotted elliptical line, the dashed elliptical line, and the
central star symbol denote the 1$\sigma$ confidence region, the
2$\sigma$ confidence region, and the best-fit point of the
I$\Lambda$CDM1 model in the $\Omega_{m}$-$h$ plane, respectively. In
the left panel, three solid diagonal lines, from top to bottom,
represent the constraints of $13.88$ Gyr, $14.02$ Gyr, and $14.2$
Gyr at $z=0$, respectively. When plotting these diagonal lines
denoting cosmic age constarints, the best-fit value of interaction
strength of I$\Lambda$CDM1 model $\alpha=-1.20\times10^{-3}$ is
adopted. Notice that the top diagonal line tangents to the 1$\sigma$
confidence region of the I$\Lambda$CDM1 model, while the middle
diagonal line tangents to the 2$\sigma$ confidence region of the
I$\Lambda$CDM1 model. In the right panel, the solid diagonal line
represents the lower limit of the quasar's age at $z=3.91$. The area
below this diagonal line corresponds to a cosmic age larger than
$1.8$ Gyr. The I$\Lambda$CDM1 model predicts a lower cosmic age than
the 1$\sigma$ lower limits of 4 GCs' age, and also predicts a lower
cosmic age than the 1$\sigma$ lower limit of the old quasar's age.}
\end{figure}

We plot Fig.\ref{fig3} to demonstrate the cosmic age problem in the I$\Lambda$CDM1 model.
The left panel of Fig.\ref{fig3} represents the cosmic age problem at $z=0$.
From this panel,
it is found that in the I$\Lambda$CDM1 model, the present cosmic age $t_0\leq 14.02$ Gyr at 2$\sigma$ CL.
Although the I$\Lambda$CDM1 model can give an upper limit of the present cosmic age 0.14 Gyr larger than that given by the $\Lambda$CDM model,
this cosmic age upper limit is still smaller than the 1$\sigma$ lower limits of 4 GCs' age
(including B024 with a lower limit 14.50 Gyr, B050 with a lower limit 15.70 Gyr, B129 with a lower limit 14.40 Gyr, and B297D with a lower limit 14.33 Gyr).
So the I$\Lambda$CDM1 model still cannot pass the cosmic age test of these 4 GCs.
We also check what the discrepancy is when all 139 GCs are taken into account.
It is found that the I$\Lambda$CDM1 model can accommodate 127 GCs,
and the statistical discrepancy disappears when all objects are included.
Besides, the right panel of Fig.\ref{fig3} represents the high-$z$ cosmic age problem at $z=3.91$.
Our calculations show that the existence of high-$z$ quasar APM 08279 + 5255 cannot be accommodated in the I$\Lambda$CDM1 model,
and the discrepancy is at 5$\sigma$ CL.

\begin{figure}
\includegraphics[scale=0.6, angle=0]{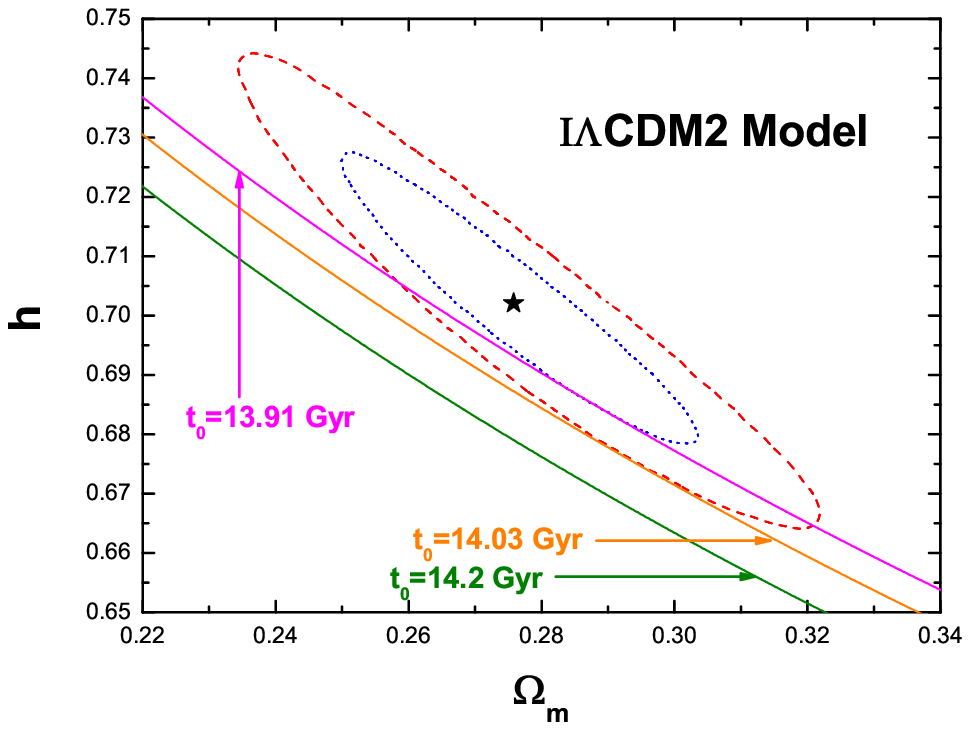}
\includegraphics[scale=0.6, angle=0]{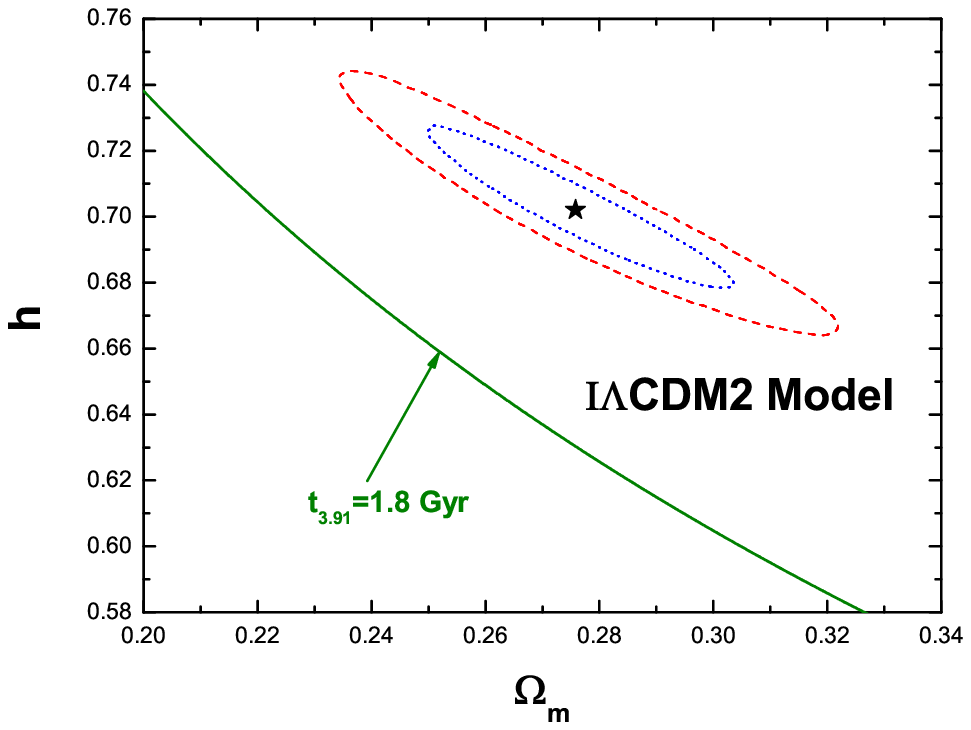}
\caption{\label{fig4} Cosmic age problem in the I$\Lambda$CDM2
model. The dotted elliptical line, the dashed elliptical line, and
the central star symbol denote the 1$\sigma$ confidence region, the
2$\sigma$ confidence region, and the best-fit point of the
I$\Lambda$CDM2 model in the $\Omega_{m}$-$h$ plane, respectively. In
the left panel, three solid diagonal lines, from top to bottom,
represent the constraints of $13.91$ Gyr, $14.03$ Gyr, and $14.2$
Gyr at $z=0$, respectively. When plotting these diagonal lines
denoting cosmic age constraints, the best-fit value of interaction
strength of the I$\Lambda$CDM2 model $\alpha=-1.43\times10^{-3}$ is
adopted. Notice that the top diagonal line tangents to the 1$\sigma$
confidence region of the I$\Lambda$CDM2 model, while the middle
diagonal line tangents to the 2$\sigma$ confidence region of the
I$\Lambda$CDM2 model. In the right panel, the solid diagonal line
represents the lower limit of the quasar's age at $z=3.91$. The area
below this diagonal line corresponds to a cosmic age larger than
$1.8$ Gyr. The I$\Lambda$CDM2 model predicts a lower cosmic age than
the 1$\sigma$ lower limits of 4 GCs' age, and also predicts a lower
cosmic age than the 1$\sigma$ lower limit of the old quasar's age.}
\end{figure}

The situation of the I$\Lambda$CDM2 model is demonstrated in the Fig.\ref{fig4}.
The left panel of Fig.\ref{fig4} represents the cosmic age problem at $z=0$.
From this panel,
it is found that in the I$\Lambda$CDM2 model, the present cosmic age $t_0\leq 14.03$ Gyr at 2$\sigma$ CL.
Although the I$\Lambda$CDM2 model can give an upper limit of the present cosmic age 0.15 Gyr larger than that given by the $\Lambda$CDM model,
this cosmic age upper limit is still smaller than the 1$\sigma$ lower limits of 4 GCs' age (including B024, B050, B129, and B297D).
We also check what the discrepancy is when all 139 GCs are taken into account.
It is found that the I$\Lambda$CDM2 model can accommodate 127 GCs,
and the statistical discrepancy disappears when all objects are included.
Besides, the right panel of Fig.\ref{fig4} represents the high-$z$ cosmic age problem at $z=3.91$.
Our calculations show that the I$\Lambda$CDM2 model cannot pass the cosmic age test of the quasar APM 08279 + 5255,
and the discrepancy is at 6$\sigma$ CL.

\begin{figure}
\includegraphics[scale=0.6, angle=0]{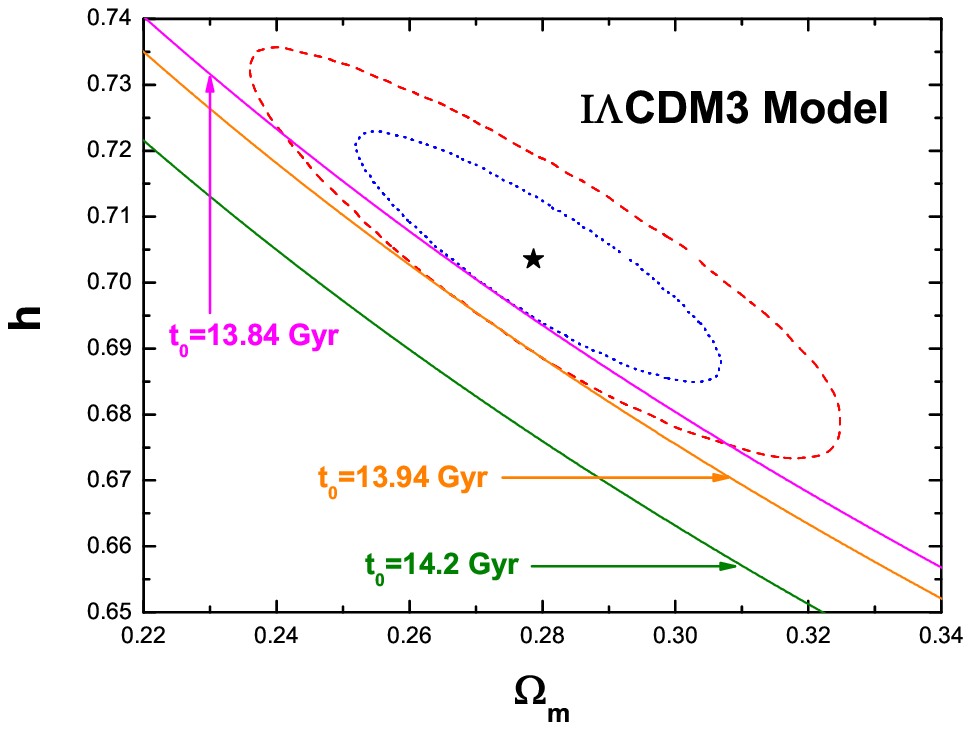}
\includegraphics[scale=0.6, angle=0]{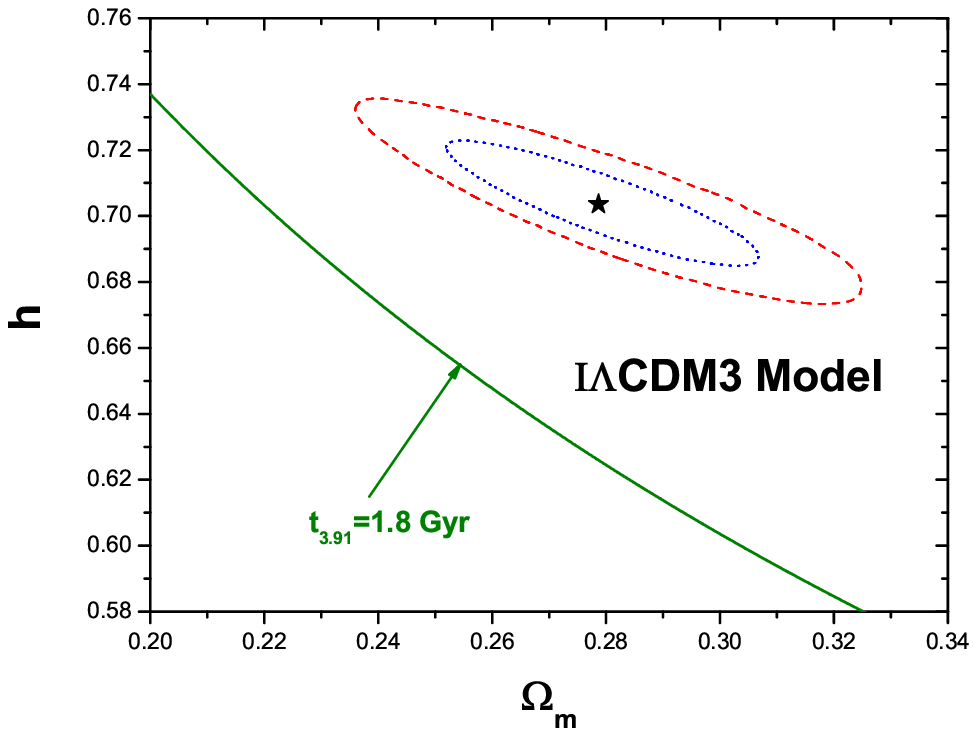}
\caption{\label{fig5} Cosmic age problem in the I$\Lambda$CDM3
model. The dotted elliptical line, the dashed elliptical line, and
the central star symbol denote the 1$\sigma$ confidence region, the
2$\sigma$ confidence region, and the best-fit point of the
I$\Lambda$CDM3 model in the $\Omega_{m}$-$h$ plane, respectively. In
the left panel, three solid diagonal lines, from top to bottom,
represent the constraints of $13.84$ Gyr, $13.94$ Gyr, and $14.2$
Gyr at $z=0$, respectively. When plotting those diagonal lines
denoting cosmic age constraints, the best-fit value of interaction
strength of the I$\Lambda$CDM3 model $\alpha=-8.98\times10^{-4}$ is
adopted. Notice that the top diagonal line tangents to the 1$\sigma$
confidence region of the I$\Lambda$CDM3 model, while the middle
diagonal line tangents to the 2$\sigma$ confidence region of the
I$\Lambda$CDM3 model. In the right panel, the solid diagonal line
represents the lower limit of the quasar's age at $z=3.91$. The area
below this diagonal line corresponds to a cosmic age larger than
$1.8$ Gyr. The I$\Lambda$CDM3 model predicts a lower cosmic age than
the 1$\sigma$ lower limits of 5 GCs' age, and also predicts a lower
cosmic age than the 1$\sigma$ lower limit of the old quasar's age.}
\end{figure}

The situation of the I$\Lambda$CDM3 model is shown in the Fig.\ref{fig5}.
The left panel of Fig.\ref{fig5} represents the cosmic age problem at $z=0$.
From this panel,
it is found that in the I$\Lambda$CDM3 model, the present cosmic age $t_0\leq 13.94$ Gyr at 2$\sigma$ CL.
Although the I$\Lambda$CDM3 model can give an upper limit of the present cosmic age larger than that given by the $\Lambda$CDM model,
this cosmic age upper limit is still smaller than the 1$\sigma$ lower limits of 5 GCs' age.
We also check what the discrepancy is when all 139 GCs are taken into account.
It is found that the I$\Lambda$CDM3 model can accommodate 127 GCs,
and the statistical discrepancy disappears when all objects are included.
Besides, the right panel of Fig.\ref{fig5} represents the high-$z$ cosmic age problem at $z=3.91$.
Our calculations show that the existence of high-$z$ quasar APM 08279 + 5255 cannot be accommodated in the I$\Lambda$CDM3 model,
and the discrepancy is at 6$\sigma$ CL.

Therefore, although the introduction of the interaction between dark
sectors can give a larger cosmic age, the interacting dark energy
models still have difficulty to pass the cosmic age test. This
conclusion is different from that of \cite{swang1}.

\

\section{Summary}\label{sec:Summary}

The cosmic age problem is a longstanding issue, and provides an
important tool for constraining the expanding history of the
universe \cite{Lan}. In this work, we investigate the cosmic age
problem associated with 9 extremely old globular clusters in M31
galaxy and 1 very old high-$z$ quasar APM 08279 + 5255 at $z=3.91$.
It should be pointed out that these 9 GCs have not been used to
study the cosmic age problem in the previous literature. By
evaluating the age of the universe in the $\Lambda$CDM model with
the constraints from the SNIa, the BAO, the CMB, and the independent
$H_0$ measurements, we find that the existence of 5 globular
clusters and 1 high-$z$ quasar are in tension (over 2$\sigma$
confidence level) with the current cosmological observations.
Therefore, if the age estimates of these objects are correct, the
cosmic age puzzle still remains in the standard cosmology. Moreover,
we extend our investigations to the cases of the interacting dark
energy models. By studying the cosmic age problem in the three classes of
interacting models, it is found that although the introduction of
the interaction between dark sectors can give a larger cosmic age,
the interacting DE models still have difficulty to pass the cosmic
age test. This conclusion is different from that of \cite{swang1}.
In a latest paper \cite{Cui10}, Cui and Zhang argue that the
high-$z$ cosmic age problem can be greatly alleviated when the
interaction and spatial curvature are both introduced in the
holographic dark energy model \cite{HDE1,HDE2}. Therefore, the
cosmic age problem still needs to be further investigated in the
future work.

\

\section*{Acknowledgements}
We are grateful to the referee for helpful suggestions.
We would like to thank Jun Ma and Deepak Jain, for helpful discussions.
This work was supported by the NSFC grant No.10535060/A050207, a NSFC group grant No.10821504
and Ministry of Science and Technology 973 program under grant No.2007CB815401.
Shuang Wang was also supported by a graduate fund of USTC.

\

\begin{center}
{\bf Appendix A: Utilizing Various Cosmological Observations to
constrain DE Models}
\end{center}

First we start with the SNIa observations. We use the Constitution
sample including 397 data that are given in terms of the distance
modulus $\mu_{obs}(z_i)$ \cite{SN09}. The theoretical distance
modulus is defined as
\begin{equation}
\mu_{th}(z_i)\equiv 5 \log_{10} {D_L(z_i)} +\mu_0,
\end{equation}
where $\mu_0\equiv 42.38-5\log_{10}h$ with $h$ the Hubble constant
$H_0$ in units of 100 km/s/Mpc, and in a flat universe the
Hubble-free luminosity distance $D_L\equiv H_0 d_L$ ($d_L$ denotes
the physical luminosity distance) is
\begin{equation}
D_L(z)=(1+z)\int_0^z {dz'\over E(z'; \theta)},
\end{equation}
where $\theta$ denotes the model parameters. The $\chi^2$ for the
SNIa data is
\begin{equation}
\chi^2_{SN}(\theta)=\sum\limits_{i=1}^{397}{[\mu_{obs}(z_i)-\mu_{th}(z_i;\theta)]^2\over
\sigma_i^2},\label{ochisn}
\end{equation}
where $\mu_{obs}(z_i)$ and $\sigma_i$ are the observed value and the
corresponding 1$\sigma$ error of distance modulus for each
supernova, respectively. Following Ref.\cite{Nesseris:2005ur}, the
minimization with respect to $\mu_0$ can be made trivially by
expanding the $\chi^2$ of Eq. (\ref{ochisn}) with respect to $\mu_0$
as
\begin{equation}
\chi^2_{SN}(\theta)=A(\theta)-2\mu_0 B(\theta)+\mu_0^2 C,
\end{equation}
where
\begin{equation}
A(\theta)=\sum\limits_{i}{[\mu_{obs}(z_i)-\mu_{th}(z_i;\mu_0=0,\theta)]^2\over
\sigma_i^2},
\end{equation}
\begin{equation}
B(\theta)=\sum\limits_{i}{\mu_{obs}(z_i)-\mu_{th}(z_i;\mu_0=0,\theta)\over
\sigma_i^2},
\end{equation}
\begin{equation}
C=\sum\limits_{i}{1\over \sigma_i^2}.
\end{equation}
Evidently, Eq.(\ref{ochisn}) has a minimum for $\mu_0=B/C$ at
\begin{equation} \label{tchi2sn}
\tilde{\chi}^2_{SN}(\theta)=A(\theta)-{B(\theta)^2\over C}.
\end{equation}
Since $\chi^2_{SN, min}=\tilde{\chi}^2_{SN,min}$, instead minimizing
$\chi^2_{SN}$ we minimize $\tilde{\chi}^2_{SN}$ which is independent
of the nuisance parameter $\mu_0$.

Next we turn to the BAO observations. The spherical average gives us
the following effective distance measure \cite{bao05}
\begin{equation}
 D_V(z) \equiv \left[(1+z)^2D_A^2(z)\frac{z}{H(z)}\right]^{1/3},
\end{equation}
where $D_A(z)$ is the proper angular diameter distance
\begin{equation}
D_A(z)=\frac{1}{1+z}\int^z_0\frac{dz^\prime}{E(z^\prime)}
\label{eq:da}
\end{equation}
The BAO data from the spectroscopic SDSS DR7 galaxy sample
\cite{BAO2} give $D_V(z=0.35)/D_V(z=0.2)=1.736\pm 0.065$. Thus, the
$\chi^2$ for the BAO data is,
\begin{equation}
\chi^2_{BAO}=\left(\frac{D_V(z=0.35)/D_V(z=0.2)-1.736}{0.065}\right)^2
\end{equation}

Next consider the CMB observations. Here we employ the ``WMAP
distance priors'' given by the 7-year WMAP observations
\cite{WMAP7}. This includes the ``acoustic scale'' $l_A$, the
``shift parameter'' $R$, and the redshift of the decoupling epoch of
photons $z_*$. The acoustic scale $l_A$ describes the distance ratio
$D_A(z_*)/r_s(z_*)$, defined as
\begin{equation}
\label{ladefeq} l_A\equiv (1+z_*){\pi D_A(z_*)\over r_s(z_*)},
\end{equation}
where a factor of $(1+z_*)$ arises because $D_A(z_*)$ is the proper
angular diameter distance, whereas $r_s(z_*)$ is the comoving sound
horizon at $z_*$. The fitting formula of $r_s(z)$ is given by
\begin{equation}
r_s(z)=\frac{1} {\sqrt{3}}  \int_0^{1/(1+z)}  \frac{ da } { a^2H(a)
\sqrt{1+(3\Omega_{b}/4\Omega_{\gamma})a} },
\end{equation}
where $\Omega_{b}$ and $\Omega_{\gamma}$ are the present-day baryon
and photon density parameters, respectively. In this paper, we adopt
the best-fit values, $\Omega_{b}=0.022765 h^{-2}$ and
$\Omega_{\gamma}=2.469\times10^{-5}h^{-2}$ (for $T_{cmb}=2.725$ K),
given by the 7-year WMAP observations \cite{WMAP7}. The fitting
function of $z_*$ is proposed by Hu and Sugiyama \cite{Hu:1995en}:
\begin{equation}
\label{zstareq} z_*=1048[1+0.00124(\Omega_b
h^2)^{-0.738}][1+g_1(\Omega_m h^2)^{g_2}],
\end{equation}
where
\begin{equation}
g_1=\frac{0.0783(\Omega_b h^2)^{-0.238}}{1+39.5(\Omega_b
h^2)^{0.763}},\quad g_2=\frac{0.560}{1+21.1(\Omega_b h^2)^{1.81}}.
\end{equation}
The shift parameter $R$ is responsible for the distance ratio
$D_A(z_*)/H^{-1}(z_*)$, given by \cite{Bond97}
\begin{equation}
\label{shift} R(z_*)\equiv \sqrt{\Omega_m H_0^2}(1+z_*)D_A(z_*).
\end{equation}
Following Ref.\cite{WMAP7}, we use the prescription for using the
WMAP distance priors. Thus, the $\chi^2$ for the CMB data is
\begin{equation}
\chi_{CMB}^2=(x^{th}_i-x^{obs}_i)(C^{-1})_{ij}(x^{th}_j-x^{obs}_j),\label{chicmb}
\end{equation}
where $x_i=(l_A, R, z_*)$ is a vector, and $(C^{-1})_{ij}$ is the
inverse covariance matrix. The 7-year WMAP observations \cite{WMAP7}
give the maximum likelihood values: $l_A(z_*)=302.09$,
$R(z_*)=1.725$, and $z_*=1091.3$. The inverse covariance matrix is
also given in Ref.~\cite{WMAP7}
\begin{equation}
(C^{-1})=\left(
  \begin{array}{ccc}
    2.305 & 29.698 & -1.333 \\
    29.698 & 6825.27 & -113.180 \\
    -1.333 & -113.180  &  3.414 \\
  \end{array}
\right).
\end{equation}

At last, we also use the prior on the present-day Hubble constant
$H_0=74.2\pm 3.6$km/s/Mpc \cite{H0}. In Ref.\cite{H0}, the authors
obtain this measured value of $H_0$ from the magnitude-redshift
relation of 240 low-$z$ type Ia supernovae at $z<0.1$. It is
remarkable that this Gaussian prior on $H_0$ has also been used in
the analysis of the WMAP 7-year observational data \cite{WMAP7}. The
$\chi^2$ function for the Hubble constant is
\begin{equation}
\chi^2_{h}=\left(\frac{h-0.742}{0.036} \right)^2.
\end{equation}


\end{document}